\documentclass[10pt,twocolumn,letterpaper]{article}

\usepackage{cvpr}              

\usepackage{adjustbox}
\usepackage{multirow}

\definecolor{cvprblue}{rgb}{0.21,0.49,0.74}
\usepackage[pagebackref,breaklinks,colorlinks,allcolors=cvprblue]{hyperref}  

\title{CADD: A Chinese Traffic Accident Dataset for Statute-Based Liability Attribution}

\author{
Yunfei Shen \\
University of Science and Technology of China (USTC)
\and
Zhongcheng Wu \\
Hefei Institutes of Physical Science, Chinese Academy of Sciences \\
and University of Science and Technology of China (USTC)
}

\begin{document}
\maketitle
\begin{abstract}
As autonomous driving technology advances, the critical challenge evolves beyond collision avoidance to the \textbf{adjudication of liability} when accidents occur. Existing datasets, focused on detection and localization, lack the annotations required for this legal reasoning. To bridge this gap, we introduce the \textbf{C}hinese \textbf{A}ccident \textbf{D}uty-determination \textbf{D}ataset (\textbf{CADD}), the first benchmark for statute-based liability attribution. CADD contains 792 real-world driving recorder videos, each annotated within a novel \textbf{``Behavior--Liability--Statute''} pipeline. This framework provides \textbf{granular, symmetric behavior annotations}, clear responsibility assignments, and, uniquely, links each case to the specific \textbf{Chinese traffic law statute} violated. We demonstrate the utility of CADD through detailed analysis and establish benchmarks for liability prediction and explainable decision-making. By directly connecting perceptual data to legal consequences, CADD provides a foundational resource for developing accountable and legally-grounded autonomous systems.
\end{abstract}    
\section{Introduction}

As autonomous driving systems approach human-level perception and decision-making, the next frontier lies in understanding not only \textit{how} accidents happen but also \textit{why} they occur and \textit{who} is responsible. Beyond safe navigation, real-world deployment demands that vehicles interpret behaviors within the framework of traffic laws, attribute liability, and justify their actions accordingly. Such capability is central to legal compliance, insurance adjudication, and—most importantly—the cultivation of public trust in autonomous systems.

Existing datasets have fueled remarkable progress in accident analysis. Benchmarks such as DoTA~\cite{DoTA} and A3D~\cite{A3D} focus on temporal and spatial understanding, effectively answering the questions of “when,” “where,” and “what” in traffic incidents. Later works like CTA~\cite{CTA} explore causal relationships between events, while SUTD-TrafficQA~\cite{SUTD-TrafficQA} introduces reasoning tasks through natural-language queries. Despite their sophistication, these datasets remain confined to perceptual and descriptive domains. They reveal what happened, but not who is legally at fault or under which statute responsibility should be assigned—questions that are indispensable for deploying autonomous vehicles in legally governed environments.

To address this gap, we introduce the \textbf{Chinese Accident Duty-determination Dataset (CADD)}, the first benchmark that bridges behavioral understanding with statute-based liability attribution. CADD provides the complete “Behavior–Liability–Statute” chain required for legally grounded decision-making. It contains 792 real-world traffic collision videos with hierarchical, symmetric annotations of all involved parties, definitive liability judgments, and explicit mappings to corresponding Chinese traffic laws.

We make three key contributions. 
First, we release a large-scale, multi-modal dataset that systematically links observable driving behaviors to legally justified outcomes. 
Second, we propose a unified “Behavior–Liability–Statute” annotation framework that captures the full reasoning process from perception to adjudication, establishing a foundation for research on legally aware autonomous systems. 
Third, we design benchmark tasks and evaluation protocols that enable exploration of automated liability attribution, causal reconstruction, and explainable legal reasoning.

CADD thus moves beyond accident detection toward comprehensive accident adjudication, paving the way for autonomous vehicles that can reason, decide, and explain their actions within the boundaries of law.

\section{Related Work}
\label{sec:related}

\subsection{Traffic Accident Datasets}

Research in traffic accident analysis has produced datasets that progressively address deeper aspects of accident understanding, yet systematically stop short of legal adjudication.

Early benchmarks like DoTA~\cite{DoTA} and A3D~\cite{A3D} established foundations for accident perception, providing temporal localization and classification of anomalies. However, their annotations remain descriptive, answering \textit{when} and \textit{what} without addressing interactive causality or liability. Subsequent datasets introduced causal reasoning: CTA~\cite{CTA} decomposes accidents into semantic "cause" and "effect" events, while SUTD-TrafficQA~\cite{SUTD-TrafficQA} explores fault concepts through visual question-answering. Nevertheless, they lack formal, instance-level liability assignments—CTA describes pre-crash behaviors without designating responsibility, while SUTD-TrafficQA's fault questions lack definitive ground-truth labels.

This limitation persists across the field: DADA2000~\cite{DADA} annotates causes but not liability; the CarCrashDataset (CCD)~\cite{CCD} explicitly notes detailed liability annotations are for future release; and even CycleCrash~\cite{CycleCrash}, which includes 'fault' annotation, omits links to specific traffic statutes. Internationally, large-scale autonomous driving datasets such as nuScenes~\cite{nuscenes}, and Lyft Level 5~\cite{Lyft} provide rich perception data and rare accident scenarios, but they similarly lack legally-grounded liability labels, highlighting a universal gap in accident adjudication research.

As summarized in Table~\ref{tab:dataset_comparison_with_size}, existing datasets universally lack the complete "Behavior–Liability–Statute" pipeline. CADD directly addresses this gap by providing the first benchmark that connects observable interactions to legal outcomes through explicit statute references.

\subsection{Driving Behavior and Risk Assessment}

Parallel research focuses on proactive safety through behavior analysis and risk prediction. Large-scale datasets like BDD100K~\cite{bdd100k} provide foundational resources for scene understanding and trajectory analysis. Finer-grained behavioral datasets include the 100-Driver dataset~\cite{100-driver} for driver distraction classification and the DAD dataset~\cite{DAD} for anomalous driving action detection. Beyond behavioral classification, significant work aims to predict accident risk through multi-source integration and video-based anticipation (e.g., AccNet~\cite{liao2024real}, SSC-TAD~\cite{fang2022traffic}, HF2-VADAD~\cite{bogdoll2024hybrid}).

While invaluable for proactive safety, these approaches typically analyze single-actor behavior or scene-level risk without modeling the symmetrical interactions essential for fault determination. CADD bridges this gap by explicitly focusing on paired vehicle interactions and mapping them directly to legal outcomes.

\subsection{Legal Reasoning and Autonomous Vehicle Liability}

Beyond perception and risk prediction, there is a growing interest in legal reasoning for autonomous vehicles. Several studies explore frameworks for assigning responsibility in AVs. For example, \cite{on_av_traffic_law} discusses how autonomous vehicles can conform to traffic laws, highlighting challenges such as the open-texture, ambiguity, and exceptions present in legal regulations. Similarly, \cite{responsibility_autonomous_systems} investigates the evolving notion of responsibility for autonomous systems, emphasizing the gap between traditional liability frameworks and AI agency.

Some works attempt to integrate traffic rules into AV decision-making using formal logic or simulation. For instance, \cite{queensland_rule_logic} employs defeasible deontic logic to model overtaking rules for decision-making, while \cite{liability_game_theory} uses a hierarchical game-theoretic approach to study liability allocation between AVs and human drivers. More recently, \cite{llm_rule_reasoning} proposes using retrieval-augmented reasoning with large language models to interpret traffic regulations for AV decision support.

While these approaches contribute valuable insights, they mainly focus on rule compliance or theoretical responsibility frameworks. Critically, existing methods rarely provide datasets that link observable vehicle behaviors to instance-level legal statutes, leaving a gap between theoretical legal reasoning and practical accident adjudication. CADD addresses this gap by offering real-world collision videos annotated with symmetric vehicle behaviors, liability determinations, and the specific traffic laws violated, establishing the first dataset suitable for legally-grounded AV research.

CADD complements these approaches by providing real-world evidence with hierarchical, multi-modal annotations grounded in legal statutes. It uniquely supports research on automated liability attribution, causal reasoning, and explainable decision-making, advancing AV development from statistically safe operation toward legally accountable behavior.

\begin{table*}[t]
\centering
\small
\caption{Comprehensive comparison of traffic datasets. Abbreviations: \#Vid (number of videos), Dur (duration in hours), TL (temporal localization), SB (symmetric behavior), LA (liability attribution), LS (legal statutes). $\checkmark$: full, $\sim$: partial, ---: none.}
\begin{tabular}{@{}lrrcccccc@{}}
\toprule
Dataset & Year & \#Vid & Dur & TL & SB & LA & LS \\
\midrule
KITTI~\cite{KITTI} & 2012 & 22 & 1.5 & --- & --- & --- & --- \\
DAD~\cite{DAD} & 2017 & 1,730 & 2.4 & $\checkmark$ & --- & --- & --- \\
CADP~\cite{CADP} & 2018 & 1,416 & 5.2 & $\checkmark$ & --- & --- & --- \\
A3D~\cite{A3D} & 2019 & 1,500 & 3.6 & $\checkmark$ & --- & --- & --- \\
BDD100K~\cite{bdd100k} & 2020 & 100k* & 1,100* & --- & --- & --- & --- \\
DoTA~\cite{DoTA} & 2020 & 4,677 & 20.0 & $\checkmark$ & --- & --- & --- \\
CTA~\cite{CTA} & 2020 & 3,000 & 3.0 & $\checkmark$ & --- & --- & --- \\
DADA-2000~\cite{DADA} & 2019 & 2,000 & 6.1 & $\checkmark$ & --- & --- & --- \\
CCD~\cite{CCD} & 2020 & 1,500 & 6.3 & $\checkmark$ & --- & --- & --- \\
SUTD-TrafficQA~\cite{SUTD-TrafficQA} & 2021 & 10,000 & --- & $\checkmark$ & --- & $\sim$ & --- \\
CycleCrash~\cite{CycleCrash} & 2024 & 3,000 & 4.1 & $\checkmark$ & --- & --- & --- \\
\hline
\textbf{CADD (Ours)} & \textbf{2025} & \textbf{792} & \textbf{2.2} & $\checkmark$ & $\checkmark$ & $\checkmark$ & $\checkmark$ \\
\bottomrule
\end{tabular}
\label{tab:dataset_comparison_with_size}
\end{table*}

\section{The Chinese Accident Duty-determination (CADD) Dataset}

\subsection{Data Collection and Curation}
The CADD dataset originates from carefully selected Chinese online platforms, including Bilibili and Weibo, where we gathered dashcam footage using traffic accident-related keywords. This sourcing strategy ensures that all scenarios operate within a consistent Chinese legal framework, providing the essential foundation for reliable liability attribution. 

To maintain high data quality and annotation integrity, we implemented a rigorous curation process. The dataset exclusively features collisions involving the ego vehicle—the dashcam-equipped vehicle—ensuring a consistent first-person perspective throughout. We focused on incidents where the responsible party could be clearly identified, deliberately filtering out ambiguous cases that might obscure the learning objective. Additionally, we excluded any footage where key participants or critical events were not fully visible. 

The final collection comprises 792 high-resolution videos totaling 134.5 minutes of footage. These clips, averaging just over 10 seconds in length, capture the decisive moments leading to and including collisions across a diverse range of driving environments and weather conditions.

\subsection{Annotation Framework}
At the heart of CADD lies its comprehensive annotation framework, which systematically structures the journey from observable events to legal conclusions. This hierarchical framework begins by capturing essential environmental context, including scene type and road geometry, then documents the behavioral dynamics through symmetric vehicle behavior annotations that record the precise actions of both ego and opponent vehicles at the critical moment. 

The framework culminates in legal outcomes, providing clear liability judgments that explicitly link to specific articles from Chinese traffic law. This creates a unique "Behavior-Liability-Statute" pipeline that directly connects observable actions to their legal consequences. The complete annotation schema, organized hierarchically from environmental context through behavioral dynamics to final outcome, is detailed in Table~\ref{tab:annotation_schema}, while the six legal statutes forming the basis for liability attribution are explicitly listed in Table~\ref{tab:legal_statutes}, ensuring full transparency and reproducibility in the legal reasoning process.

\subsection{Annotation Process and Quality Control}
To ensure the highest standard of annotation quality, particularly for the crucial liability judgments, we established a rigorous multi-stage workflow. A team of twelve annotators, all with substantial driving experience, underwent specialized training developed in consultation with traffic police to deepen their understanding of traffic laws and liability principles.

The annotation process employed a hybrid approach where some contextual labels were efficiently generated using pre-trained models and subsequently verified, while the core elements of symmetric behavior, liability assignment, and statute matching were meticulously performed by human annotators. To guarantee the reliability of these subjective judgments, we implemented a consensus-based quality assurance protocol where each video was independently evaluated by three annotators, with definitive liability labels assigned only upon full consensus. In cases of disagreement, traffic police experts provided final arbitration, ensuring the legal soundness and reliability of the dataset's most critical judgments.

\begin{table*}[t]
\centering
\caption{Annotation schema of the proposed dataset, organized by hierarchical level, classification dimension, and corresponding fields. Each field lists its category options and the total number of classes.}
\begin{tabular}{p{2cm} p{3cm} p{3cm} p{5cm} p{1cm}}
\toprule
\textbf{Level} & \textbf{Dimension} & \textbf{Field Name} & \textbf{Category Options} & \textbf{\#Classes} \\
\midrule
\multirow{2}{*}{\textbf{Environment}} 
 & Scene Type & \texttt{scene} & Highway, Urban, Rural & 3 \\
 & Road Structure & \texttt{linear} & Arterial Road, Intersection, T-junction & 3 \\
\midrule
\multirow{4}{*}{\textbf{Behavior}} 
 & Accident Type & \texttt{accident\_type} & Rear-end, Lane Change, Wrong-Way Driving, Backing, Failure to Yield (Turn), Yield to the Right & 6 \\
& Ego / Other Vehicle Behavior & \texttt{ego\_behavior}, \texttt{other\_behavior} & 
Straight, Left Lane Change, Right Lane Change, Left Turn, Right Turn, Reversing, Braking, Stationary & 8 \\

& Traffic Violation & \texttt{violation} & Refer to five legal statutes (see Table~\ref{tab:legal_statutes}) & 6 \\

\midrule
\multirow{6}{*}{\textbf{Outcome}} 
 & Collision Type & \texttt{collision\_type} & Rear-end, Side, Front, Backing & 4 \\
 & Responsibility & \texttt{responsibility} & No Fault, Full Fault & 2 \\
 & Description & \texttt{description} & Text description of accident scenario & -- \\
 & Trajectory Path & \texttt{trajectory} & All collision vehicle trajectories (frame-wise) & -- \\
 & Viewpoint & \texttt{viewpoint} & Front View, Rear Collision View & 2 \\
\bottomrule
\end{tabular}
\label{tab:annotation_schema}
\end{table*}

\begin{table*}[t]
\centering
\caption{List of legal statutes referenced for liability attribution.}
\begin{tabular}{ccc}
\toprule
\textbf{ID} & \textbf{Legal Statute} & \textbf{Description} \\
\midrule
1 & Implementation Regulation Article 52(2) & Yielding to the right at uncontrolled intersections. \\
2 & Implementation Regulation Article 52(3) & Turning vehicles must yield to straight-going vehicles. \\
3 & Road Traffic Safety Law Article 35 & Vehicles must keep to the right side of the road. \\
4 & Implementation Regulation Article 44(2) & Lane-changing vehicles must not obstruct or endanger others. \\
5 & Road Traffic Safety Law Article 43 & Drivers must maintain a safe following distance. \\
6 & Implementation Regulation Article 50 & When reversing, drivers must ensure safety before moving. \\
\bottomrule
\end{tabular}
\label{tab:legal_statutes}
\end{table*}

\subsection{Dataset Statistics and Analysis}

The statistical analysis of the CADD dataset reveals a balanced and diverse collection of traffic accident scenarios. Among the 792 cases, the ego vehicle was judged fully responsible in 209 instances (26.4\%), while being not at fault in the remaining 583 cases (73.6\%). This distribution prevents machine learning models from developing a trivial bias that the ego vehicle is always innocent, thereby encouraging more nuanced understanding of fault attribution.

\begin{figure*}[htbp]
\centering
\includegraphics[width=0.9\textwidth]{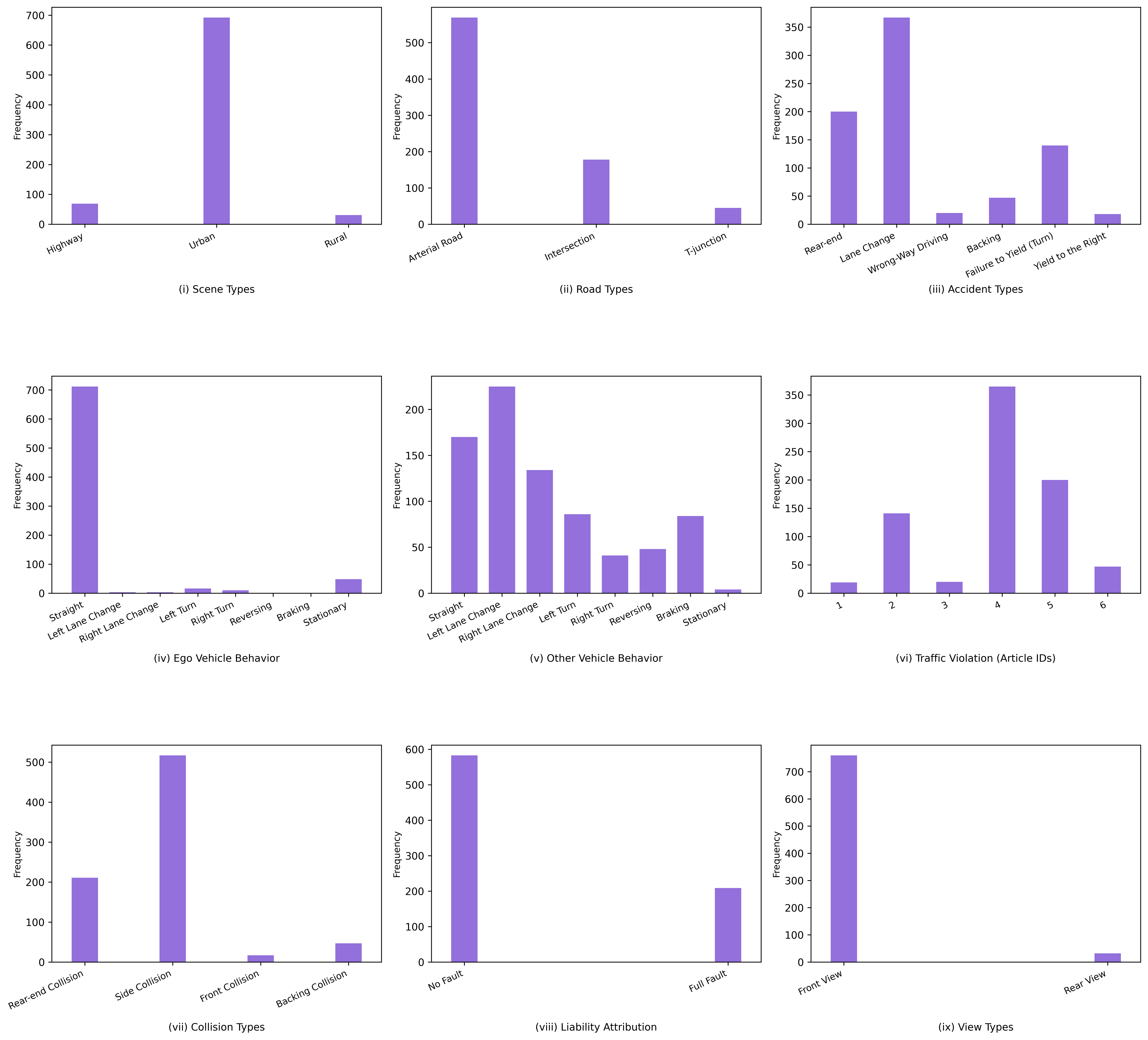}
\caption{Detailed annotation statistics of the CADD dataset across nine dimensions: (i) scene types, (ii) road types, (iii) accident types, (iv) ego vehicle behavior, (v) other vehicle behavior, (vi) traffic violation statutes, (vii) collision types, (viii) liability attribution, and (ix) view types.}
\label{fig:cadd_detailed_stats}
\end{figure*}

As illustrated in Figure~\ref{fig:cadd_detailed_stats}, the dataset exhibits rich diversity across multiple dimensions. Urban scenarios dominate the collection (87.4\%), complemented by highway (8.7\%) and rural (3.9\%) environments. The road infrastructure distribution shows arterial roads (72.0\%), intersections (22.3\%), and T-junctions (5.7\%). 

Accident type analysis reveals lane change incidents as the most frequent category (46.3\%), followed by rear-end collisions (25.3\%) and failure to yield situations (17.7\%). This distribution aligns with the collision type statistics, where side collisions predominate (65.3\%) over rear-end collisions (26.6\%). 

Behavioral annotations demonstrate distinct patterns between vehicles: the ego vehicle maintains straight trajectory in 89.9\% of cases, while other vehicles exhibit more varied maneuvers including lane changes (45.3\%) and turning actions (16.0\%). Legal statute violations are primarily attributed to lane change infractions (46.1\%) and rear-end collisions (25.3\%).

\begin{figure*}[htbp]
\centering
\includegraphics[width=0.9\textwidth]{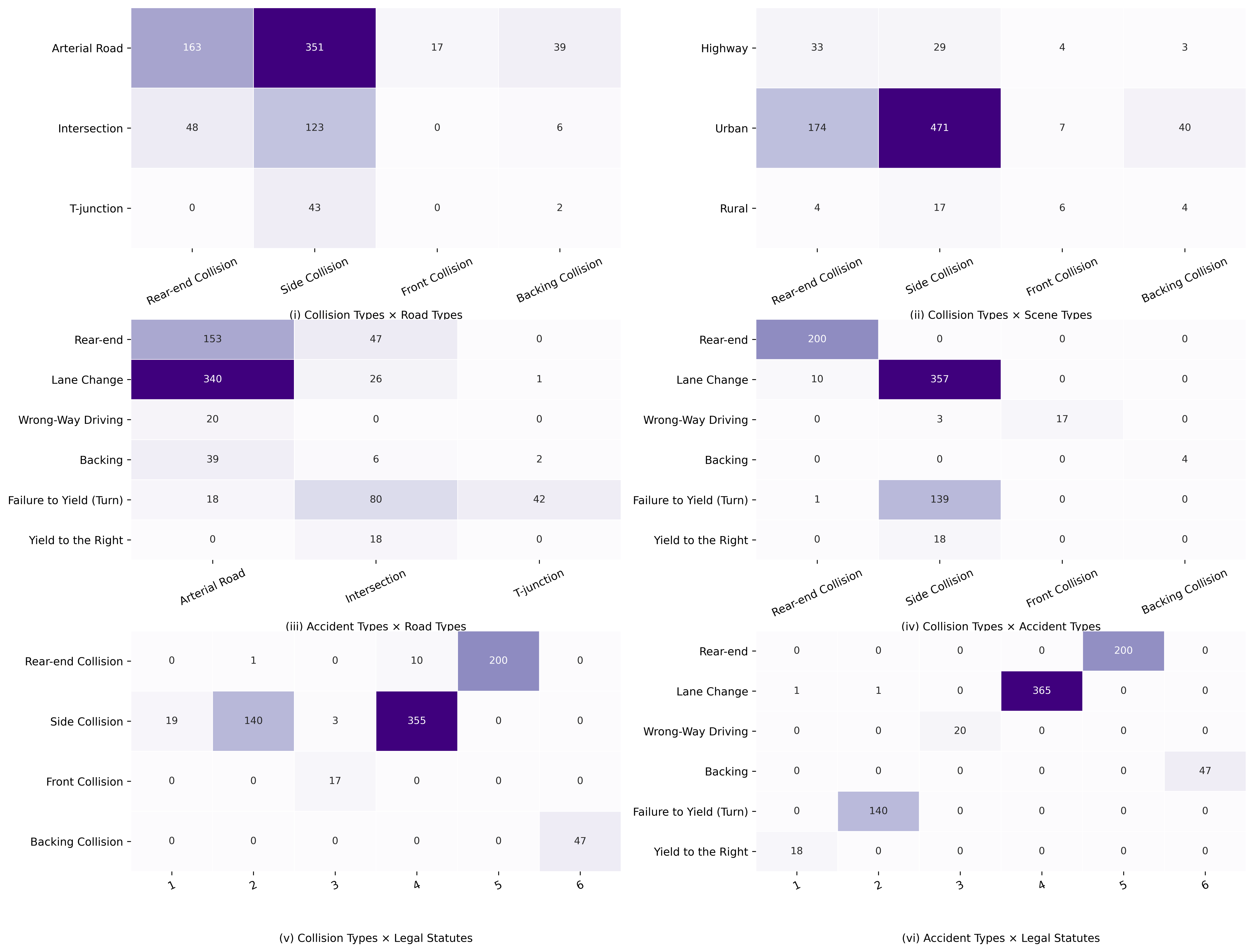}
\caption{Correlation analysis of traffic accident factors in the CADD dataset: (i) collision types × road types, (ii) collision types × scene types, (iii) accident types × road types, (iv) collision types × accident types, (v) collision types × legal statutes, and (vi) accident types × legal statutes.}
\label{fig:cadd_correlations}
\end{figure*}

To deepen our understanding of the complex relationships within traffic accidents, Figure~\ref{fig:cadd_correlations} presents six correlation heatmaps examining the interplay between collision types, road types, scene types, accident types, and legal statutes. These analyses reveal meaningful patterns that provide valuable insights for developing context-aware accident analysis systems capable of addressing the multifaceted nature of real-world traffic incidents.

The dataset's balanced liability distribution, combined with its comprehensive coverage of environmental conditions, behavioral dynamics, and legal outcomes, establishes CADD as a robust benchmark for advancing research in automated liability attribution.

\section{Data Overview and Case Studies}

To provide an intuitive understanding of the CADD dataset's composition and annotation quality, this section presents detailed case studies that exemplify the diverse scenarios captured in our collection. These examples vividly demonstrate how our "Behavior-Liability-Statute" annotation framework operates in practice across different traffic contexts.

Figure~\ref{fig:case_examples} showcases six representative accident cases that span the spectrum of scenarios in our dataset. Each case presents a composite visualization of the critical moments alongside a comprehensive annotation summary, illustrating the direct connection between observable vehicle behaviors and legal outcomes.

The selected cases encompass various common traffic scenarios: lane change accidents on arterial roads (Case 1\_008), intersection collisions involving right-of-way principles (Case 4\_003), rear-end collisions at intersections (Case 4\_009), backing incidents in urban settings (Case 11\_004), wrong-way driving on highways (Case 36\_004), and failure-to-yield situations during turning maneuvers (Case 46\_005).

Notably, these examples demonstrate the balanced responsibility distribution within our dataset, with the ego vehicle being assigned full fault in appropriate circumstances (Cases 4\_003 and 4\_009) while being exonerated in others where the other vehicle clearly violated traffic statutes. The case studies also highlight the symmetric behavior annotation approach, capturing both ego and other vehicle maneuvers with precise temporal alignment to the collision moment.

Each case explicitly links the observed traffic violation to the specific legal statute that governs the liability determination, providing transparent legal reasoning that enhances the educational and research value of the dataset. This direct mapping from behavioral observation through liability attribution to statutory foundation represents the core innovation of CADD and enables the development of legally-grounded autonomous decision systems.
\newcommand{\field}[2]{\textbf{#1:}\! #2}
\begin{figure*}[t]
  \centering

  \begin{minipage}[t]{0.48\linewidth}
    \centering
    \includegraphics[width=\linewidth]{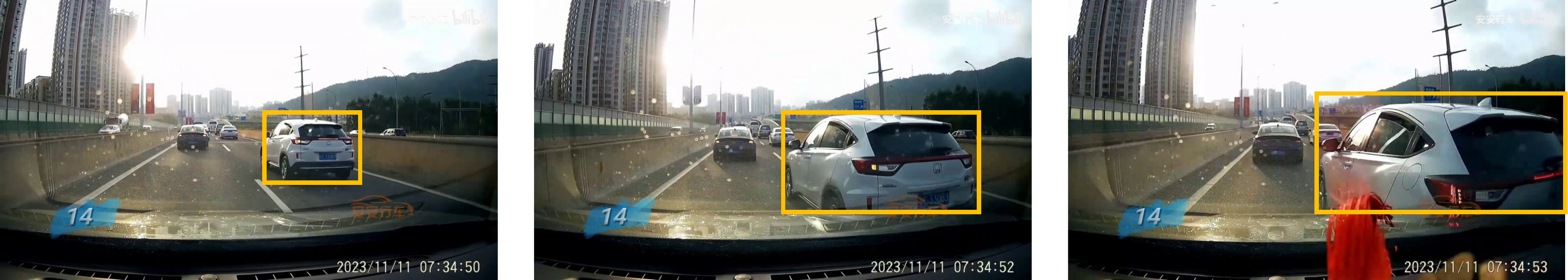}\\[2pt]
    \scriptsize
    \textbf{(a)} Case 1\_008\\[-1pt]
    \begin{tabular}{@{}ll@{\hspace{1.5em}}ll@{}}
      \field{Scene}{Urban} & & \field{Road Type}{Arterial Road} \\
      \field{Accident Type}{Lane Change} & & \field{Collision Type}{Side Collision} \\
      \field{Ego Behavior}{Straight} & & \field{Other Behavior}{Left Lane Change} \\
      \field{Responsibility}{No Fault} & & \field{Traffic Violation}{Art.~44(2)}
    \end{tabular}
  \end{minipage}
  \hfill
  \begin{minipage}[t]{0.48\linewidth}
    \centering
    \includegraphics[width=\linewidth]{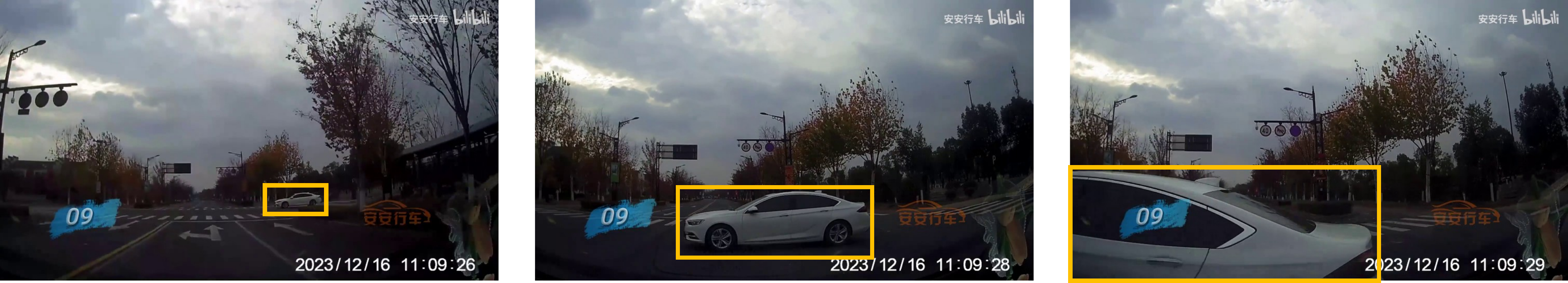}\\[2pt]
    \scriptsize
    \textbf{(b)} Case 4\_003\\[-1pt]
    \begin{tabular}{@{}ll@{\hspace{1.5em}}ll@{}}
      \field{Scene}{Urban} & & \field{Road Type}{Intersection} \\
      \field{Accident Type}{Yield to Right} & & \field{Collision Type}{Side Collision} \\
      \field{Ego Behavior}{Straight} & & \field{Other Behavior}{Straight} \\
      \field{Responsibility}{Full Fault} & & \field{Traffic Violation}{Art.~52(2)}
    \end{tabular}
  \end{minipage}

  \vspace{0.8em}

  \begin{minipage}[t]{0.48\linewidth}
    \centering
    \includegraphics[width=\linewidth]{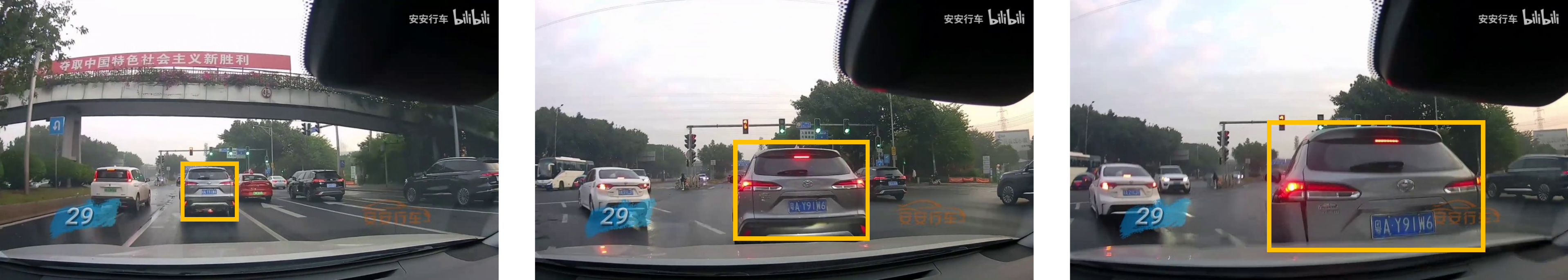}\\[2pt]
    \scriptsize
    \textbf{(c)} Case 4\_009\\[-1pt]
    \begin{tabular}{@{}ll@{\hspace{1.5em}}ll@{}}
      \field{Scene}{Urban} & & \field{Road Type}{Intersection} \\
      \field{Accident Type}{Rear-end} & & \field{Collision Type}{Rear-end Collision} \\
      \field{Ego Behavior}{Straight} & & \field{Other Behavior}{Braking} \\
      \field{Responsibility}{Full Fault} & & \field{Traffic Violation}{Art.~43}
    \end{tabular}
  \end{minipage}
  \hfill
  \begin{minipage}[t]{0.48\linewidth}
    \centering
    \includegraphics[width=\linewidth]{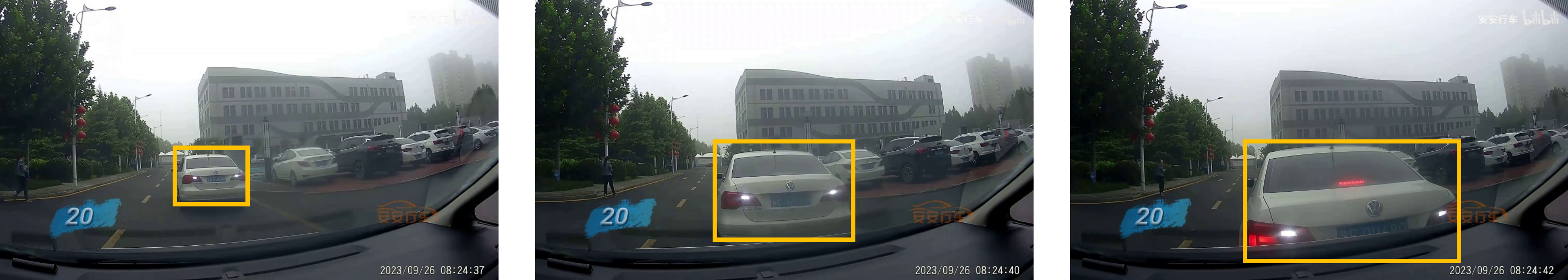}\\[2pt]
    \scriptsize
    \textbf{(d)} Case 11\_004\\[-1pt]
    \begin{tabular}{@{}ll@{\hspace{1.5em}}ll@{}}
      \field{Scene}{Urban} & & \field{Road Type}{Intersection} \\
      \field{Accident Type}{Backing} & & \field{Collision Type}{Backing Collision} \\
      \field{Ego Behavior}{Straight} & & \field{Other Behavior}{Reversing} \\
      \field{Responsibility}{No Fault} & & \field{Traffic Violation}{Art.~50}
    \end{tabular}
  \end{minipage}

  \vspace{0.8em}

  \begin{minipage}[t]{0.48\linewidth}
    \centering
    \includegraphics[width=\linewidth]{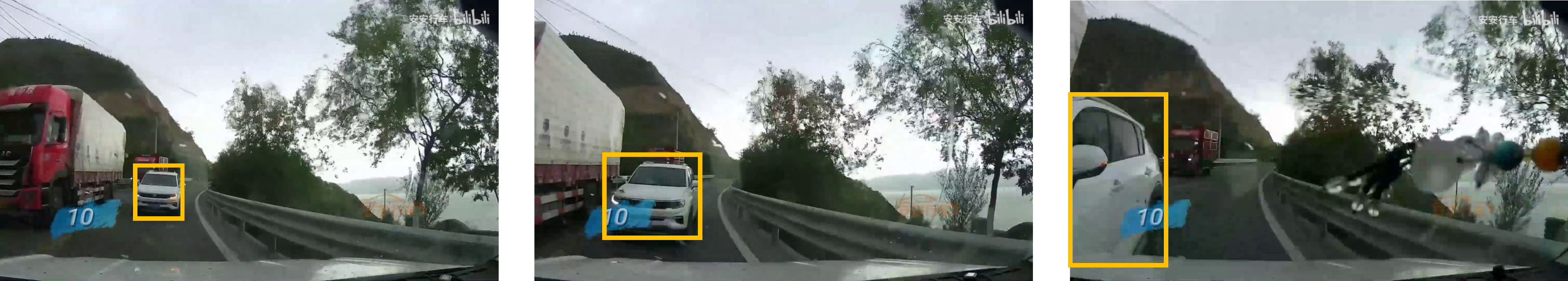}\\[2pt]
    \scriptsize
    \textbf{(e)} Case 36\_004\\[-1pt]
    \begin{tabular}{@{}ll@{\hspace{1.5em}}ll@{}}
      \field{Scene}{Highway} & & \field{Road Type}{Arterial Road} \\
      \field{Accident Type}{Wrong-Way Driving} & & \field{Collision Type}{Front Collision} \\
      \field{Ego Behavior}{Straight} & & \field{Other Behavior}{Wrong-Way Driving} \\
      \field{Responsibility}{No Fault} & & \field{Traffic Violation}{Art.~35}
    \end{tabular}
  \end{minipage}
  \hfill
  \begin{minipage}[t]{0.48\linewidth}
    \centering
    \includegraphics[width=\linewidth]{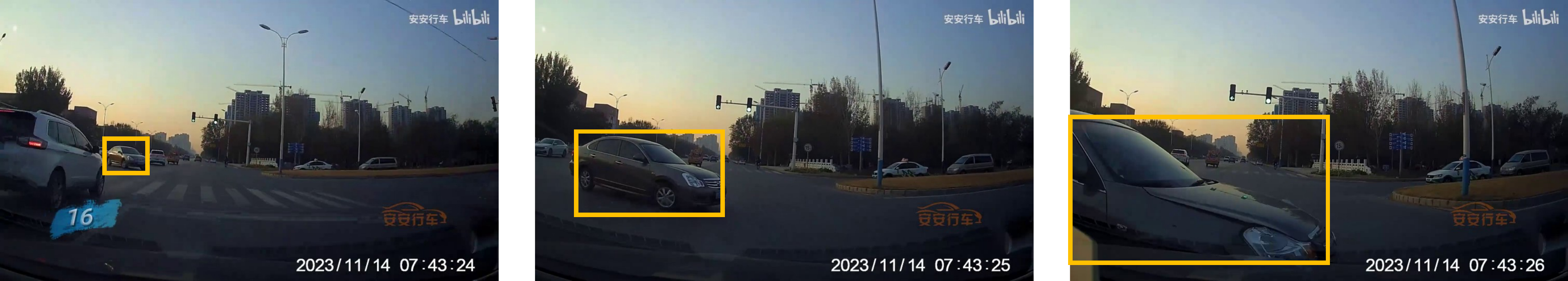}\\[2pt]
    \scriptsize
    \textbf{(f)} Case 46\_005\\[-1pt]
    \begin{tabular}{@{}ll@{\hspace{1.5em}}ll@{}}
      \field{Scene}{Urban} & & \field{Road Type}{Intersection} \\
      \field{Accident Type}{Failure to Yield (Turn)} & & \field{Collision Type}{Side Collision} \\
      \field{Ego Behavior}{Straight} & & \field{Other Behavior}{Left Turn} \\
      \field{Responsibility}{No Fault} & & \field{Traffic Violation}{Art.~52(3)}
    \end{tabular}
  \end{minipage}

  \caption{Annotated accident examples with paired image–text presentation. Each example includes a composite image (top) and compact two-column annotation summary (bottom). (a–f) represent diverse accident types across different environments, behaviors, and legal outcomes.}
  \label{fig:case_examples}
\end{figure*}

\section{Potential Applications and Benchmarking}

The CADD dataset, with its unique \emph{Behavior-Liability-Statute} annotation framework and multi-modal, frame-wise trajectory data, provides a unified foundation for studying traffic accidents at multiple levels of reasoning. Unlike traditional datasets that primarily focus on collision detection or perception tasks, CADD captures the full spectrum of traffic incidents, from the behavioral intentions of each participant to the final legal outcome. This enables models to reason about how and why an accident occurred, and under which legal principles liability should be assigned. The layered design of CADD makes it suitable not only for advancing perception-based research in autonomous driving, but also for connecting perception to decision-making, causality, and legal accountability.

\subsection{Automated Liability Attribution}

At the most fundamental level, CADD enables the development of automated liability attribution systems. Existing traffic datasets are limited to event detection and lack the relational structure required for assigning fault among multiple agents. CADD addresses this limitation through symmetric behavior annotations that record the actions of every participant involved in an accident. This allows models to analyze how each driver’s choices contribute to the outcome, shifting the focus from isolated perception to relational reasoning.

For example, in a multi-vehicle intersection collision, a model trained on CADD can evaluate whether the ego vehicle’s decision to accelerate was reasonable given the trajectory of the other vehicle, and determine liability based on codified legal rules. Such capability is particularly relevant for applications in automated insurance assessment, on-board accident adjudication, and intelligent traffic monitoring. By providing both behavioral evidence and direct mapping to legal statutes, CADD makes it possible to train systems that mirror the reasoning process of human traffic officers or legal experts.

\subsection{Legal Statute Identification}

Beyond identifying who is at fault, CADD supports the more challenging task of determining which specific laws have been violated. Each annotated case includes detailed legal references, enabling supervised learning of mappings from real-world behaviors to codified rules. This task pushes models toward a higher level of semantic understanding, recognizing not only that a vehicle ran a red light, but also that this action corresponds to a violation under a particular article of the traffic code.

This fine-grained legal reasoning is essential for interpretable and justifiable automated systems. It also opens opportunities for integrating multimodal vision-language models, where textual legal provisions guide the interpretation of visual evidence. Through this bridge between perception and law, CADD enables a new generation of AI systems that can not only make accurate judgments but also cite the statutory basis for their decisions.

\subsection{Temporal and Causal Analysis}

The frame-wise trajectory and behavior annotations in CADD enable deep temporal and causal reasoning, allowing researchers to reconstruct the full narrative of an accident rather than analyzing isolated frames. This supports models that can identify the precise sequence of interactions, such as a sudden lane change, delayed braking, or obstructed visibility, that collectively lead to a collision.

Temporal analysis also facilitates counterfactual reasoning, enabling models to simulate alternative timelines and assess whether an accident could have been avoided. For instance, a model might infer that if the ego vehicle had reduced speed half a second earlier, the collision would not have occurred. This level of understanding is crucial for developing proactive safety systems that anticipate risks rather than merely reacting to them. Methodologically, this area invites exploration of causal inference frameworks, graph-based temporal models, and transformer architectures capable of long-range reasoning across multimodal inputs.

\subsection{Explainable Decision Generation}

A major challenge in deploying autonomous systems lies in the opacity of their decision-making processes. CADD provides a foundation for developing explainable models that can articulate the rationale behind their conclusions in human-understandable and legally consistent terms. By combining video, trajectory, and statute-level annotations, the dataset enables models to translate sensory data into structured explanations, moving from technical outputs such as “the detector confidence dropped” to human-like justifications such as “the system yielded because the other vehicle had the legal right-of-way.”

Explainable decision generation is not only a matter of trust but also a prerequisite for regulatory approval and social acceptance of autonomous technologies. Research based on CADD can explore multimodal large language models or retrieval-augmented architectures that ground explanations in traffic law. In doing so, the dataset connects perception-level understanding with transparent reasoning suitable for real-world adjudication and policymaking.

\subsection{Counterfactual Safety Analysis}

At the highest level of complexity, CADD supports counterfactual safety analysis, which asks not only what happened and why, but also what could have been done differently. By leveraging detailed behavioral sequences and environmental context, models can propose alternative actions that might have prevented an accident while still adhering to traffic laws. This represents a synthesis of perception, prediction, causal reasoning, and legal awareness.

For example, a system could identify that a rear-end collision might have been avoided if the following vehicle had maintained a greater distance under rainy conditions, referencing the relevant statute governing safe following distances. This kind of reasoning supports the design of intelligent safety systems that actively learn from past incidents and contribute to adaptive driving policies and risk-aware planning.

\subsection{Progressive Benchmarking}

The diversity of annotations in CADD allows for a structured progression of benchmark tasks that correspond to increasing levels of cognitive and legal complexity. Researchers can begin with basic liability classification, move to fine-grained statute identification, then to temporal and causal reconstruction, followed by explainable decision generation, and finally to counterfactual reasoning. Each level builds upon the previous one, reflecting the hierarchical nature of human legal reasoning from factual recognition to causal understanding and normative judgment.

This progressive structure not only provides a standardized evaluation protocol but also encourages the development of models that integrate perception, reasoning, and explanation. Through this progression, CADD evolves from a static dataset into a comprehensive research framework for building AI systems capable of understanding, reasoning, and justifying their decisions within the constraints of law and human expectations.

In sum, CADD is more than a collection of annotated traffic videos. It is a comprehensive research infrastructure that connects visual perception with legal understanding. It supports a spectrum of tasks that collectively advance automated liability adjudication, traffic safety intelligence, explainable autonomous systems, and computational law. Through this integration, CADD lays the foundation for a new generation of AI that is not only accurate but also interpretable, lawful, and socially responsible.
{

{
    \small
    \begin{thebibliography}{22}
\providecommand{\natexlab}[1]{#1}
\providecommand{\url}[1]{\texttt{#1}}
\expandafter\ifx\csname urlstyle\endcsname\relax
  \providecommand{\doi}[1]{doi: #1}\else
  \providecommand{\doi}{doi: \begingroup \urlstyle{rm}\Url}\fi

\bibitem[Bao et~al.(2020)Bao, Yu, and Kong]{CCD}
Wentao Bao, Qi Yu, and Yu Kong.
\newblock Uncertainty-based traffic accident anticipation with spatio-temporal
  relational learning.
\newblock In \emph{Proceedings of the 28th ACM International Conference on
  Multimedia}, pages 2682--2690, 2020.

\bibitem[Bhuiyan et~al.(2023)Bhuiyan, Governatori, Rakotonirainy, Wong, and
  Mahajan]{queensland_rule_logic}
Hanif Bhuiyan, Guido Governatori, Andry Rakotonirainy, Meng~Weng Wong, and
  Avishkar Mahajan.
\newblock Driving decision making of autonomous vehicle according to queensland
  overtaking traffic rules.
\newblock \emph{The Review of Socionetwork Strategies}, 17\penalty0
  (2):\penalty0 233--254, 2023.

\bibitem[Bogdoll et~al.(2024)Bogdoll, Imhof, Joseph, Pavlitska, and
  Z{\"o}llner]{bogdoll2024hybrid}
Daniel Bogdoll, Jan Imhof, Tim Joseph, Svetlana Pavlitska, and J~Marius
  Z{\"o}llner.
\newblock Hybrid video anomaly detection for anomalous scenarios in autonomous
  driving.
\newblock \emph{arXiv preprint arXiv:2406.06423}, 2024.

\bibitem[Caesar et~al.(2020)Caesar, Bankiti, Lang, Vora, Liong, Xu, Krishnan,
  Pan, Baldan, and Beijbom]{nuscenes}
Holger Caesar, Varun Bankiti, Alex~H Lang, Sourabh Vora, Venice~Erin Liong,
  Qiang Xu, Anush Krishnan, Yu Pan, Giancarlo Baldan, and Oscar Beijbom.
\newblock nuscenes: A multimodal dataset for autonomous driving.
\newblock In \emph{Proceedings of the IEEE/CVF conference on computer vision
  and pattern recognition}, pages 11621--11631, 2020.

\bibitem[Cai et~al.(2024)Cai, Liu, Zhou, Ma, Zhao, Wu, and
  Ma]{llm_rule_reasoning}
Tianhui Cai, Yifan Liu, Zewei Zhou, Haoxuan Ma, Seth~Z Zhao, Zhiwen Wu, and
  Jiaqi Ma.
\newblock Driving with regulation: Interpretable decision-making for autonomous
  vehicles with retrieval-augmented reasoning via llm.
\newblock \emph{arXiv preprint arXiv:2410.04759}, 2024.

\bibitem[Chan et~al.(2016)Chan, Chen, Xiang, and Sun]{DAD}
Fu-Hsiang Chan, Yu-Ting Chen, Yu Xiang, and Min Sun.
\newblock Anticipating accidents in dashcam videos.
\newblock In \emph{Asian conference on computer vision}, pages 136--153.
  Springer, 2016.

\bibitem[Desai et~al.(2025)Desai, Etemad, and Greenspan]{CycleCrash}
Nishq~Poorav Desai, Ali Etemad, and Michael Greenspan.
\newblock Cyclecrash: A dataset of bicycle collision videos for collision
  prediction and analysis.
\newblock In \emph{2025 IEEE/CVF Winter Conference on Applications of Computer
  Vision (WACV)}, pages 6688--6698. IEEE, 2025.

\bibitem[Di et~al.(2020)Di, Chen, and Talley]{liability_game_theory}
Xuan Di, Xu Chen, and Eric Talley.
\newblock Liability design for autonomous vehicles and human-driven vehicles: A
  hierarchical game-theoretic approach.
\newblock \emph{Transportation research part C: emerging technologies},
  118:\penalty0 102710, 2020.

\bibitem[Fang et~al.(2021)Fang, Yan, Qiao, Xue, and Yu]{DADA}
Jianwu Fang, Dingxin Yan, Jiahuan Qiao, Jianru Xue, and Hongkai Yu.
\newblock Dada: Driver attention prediction in driving accident scenarios.
\newblock \emph{IEEE transactions on intelligent transportation systems},
  23\penalty0 (6):\penalty0 4959--4971, 2021.

\bibitem[Fang et~al.(2022)Fang, Qiao, Bai, Yu, and Xue]{fang2022traffic}
Jianwu Fang, Jiahuan Qiao, Jie Bai, Hongkai Yu, and Jianru Xue.
\newblock Traffic accident detection via self-supervised consistency learning
  in driving scenarios.
\newblock \emph{IEEE Transactions on Intelligent Transportation Systems},
  23\penalty0 (7):\penalty0 9601--9614, 2022.

\bibitem[Geiger et~al.(2012)Geiger, Lenz, and Urtasun]{KITTI}
Andreas Geiger, Philip Lenz, and Raquel Urtasun.
\newblock Are we ready for autonomous driving? the kitti vision benchmark
  suite.
\newblock In \emph{2012 IEEE conference on computer vision and pattern
  recognition}, pages 3354--3361. IEEE, 2012.

\bibitem[Houston et~al.(2021)Houston, Zuidhof, Bergamini, Ye, Chen, Jain,
  Omari, Iglovikov, and Ondruska]{Lyft}
John Houston, Guido Zuidhof, Luca Bergamini, Yawei Ye, Long Chen, Ashesh Jain,
  Sammy Omari, Vladimir Iglovikov, and Peter Ondruska.
\newblock One thousand and one hours: Self-driving motion prediction dataset.
\newblock In \emph{Conference on Robot Learning}, pages 409--418. PMLR, 2021.

\bibitem[Liao et~al.(2024)Liao, Li, Li, Bian, Lee, Cui, Zhang, and
  Xu]{liao2024real}
Haicheng Liao, Yongkang Li, Zhenning Li, Zilin Bian, Jaeyoung Lee, Zhiyong Cui,
  Guohui Zhang, and Chengzhong Xu.
\newblock Real-time accident anticipation for autonomous driving through
  monocular depth-enhanced 3d modeling.
\newblock \emph{Accident Analysis \& Prevention}, 207:\penalty0 107760, 2024.

\bibitem[Prakken(2017)]{on_av_traffic_law}
Henry Prakken.
\newblock On the problem of making autonomous vehicles conform to traffic law.
\newblock \emph{Artificial Intelligence and Law}, 25\penalty0 (3):\penalty0
  341--363, 2017.

\bibitem[Shah et~al.(2018)Shah, Lamare, Nguyen-Anh, and Hauptmann]{CADP}
Ankit~Parag Shah, Jean-Bapstite Lamare, Tuan Nguyen-Anh, and Alexander
  Hauptmann.
\newblock Cadp: A novel dataset for cctv traffic camera based accident
  analysis.
\newblock In \emph{2018 15th IEEE International Conference on Advanced Video
  and Signal Based Surveillance (AVSS)}, pages 1--9. IEEE, 2018.

\bibitem[Wang et~al.(2023)Wang, Li, Li, Zhang, Wu, Zhong, and Sebe]{100-driver}
Jing Wang, Wenjing Li, Fang Li, Jun Zhang, Zhongcheng Wu, Zhun Zhong, and Nicu
  Sebe.
\newblock 100-driver: A large-scale, diverse dataset for distracted driver
  classification.
\newblock \emph{IEEE Transactions on Intelligent Transportation Systems},
  24\penalty0 (7):\penalty0 7061--7072, 2023.

\bibitem[Xu et~al.(2021)Xu, Huang, and Liu]{SUTD-TrafficQA}
Li Xu, He Huang, and Jun Liu.
\newblock Sutd-trafficqa: A question answering benchmark and an efficient
  network for video reasoning over traffic events.
\newblock In \emph{Proceedings of the IEEE/CVF conference on computer vision
  and pattern recognition}, pages 9878--9888, 2021.

\bibitem[Yao et~al.(2019)Yao, Xu, Wang, Crandall, and Atkins]{A3D}
Yu Yao, Mingze Xu, Yuchen Wang, David~J Crandall, and Ella~M Atkins.
\newblock Unsupervised traffic accident detection in first-person videos.
\newblock In \emph{2019 IEEE/RSJ International conference on intelligent robots
  and systems (IROS)}, pages 273--280. IEEE, 2019.

\bibitem[Yao et~al.(2020)Yao, Wang, Xu, Pu, Atkins, and Crandall]{DoTA}
Yu Yao, Xizi Wang, Mingze Xu, Zelin Pu, Ella Atkins, and David Crandall.
\newblock When, where, and what? a new dataset for anomaly detection in driving
  videos.
\newblock \emph{arXiv preprint arXiv:2004.03044}, 2020.

\bibitem[Yazdanpanah et~al.(2023)Yazdanpanah, Gerding, Stein, Dastani, Jonker,
  Norman, and Ramchurn]{responsibility_autonomous_systems}
Vahid Yazdanpanah, Enrico~H Gerding, Sebastian Stein, Mehdi Dastani,
  Catholijn~M Jonker, Timothy~J Norman, and Sarvapali~D Ramchurn.
\newblock Reasoning about responsibility in autonomous systems: challenges and
  opportunities.
\newblock \emph{Ai \& Society}, 38\penalty0 (4):\penalty0 1453--1464, 2023.

\bibitem[You and Han(2020)]{CTA}
Tackgeun You and Bohyung Han.
\newblock Traffic accident benchmark for causality recognition.
\newblock In \emph{European Conference on Computer Vision}, pages 540--556.
  Springer, 2020.

\bibitem[Yu et~al.(2020)Yu, Chen, Wang, Xian, Chen, Liu, Madhavan, and
  Darrell]{bdd100k}
Fisher Yu, Haofeng Chen, Xin Wang, Wenqi Xian, Yingying Chen, Fangchen Liu,
  Vashisht Madhavan, and Trevor Darrell.
\newblock Bdd100k: A diverse driving dataset for heterogeneous multitask
  learning.
\newblock In \emph{Proceedings of the IEEE/CVF conference on computer vision
  and pattern recognition}, pages 2636--2645, 2020.

\end{thebibliography}
  
}
    
}


\end{document}